\documentclass[twocolumn]{aastex63} 

\usepackage{graphicx} 
\usepackage{float} 
\usepackage{wrapfig} 
\usepackage{lipsum} 
\usepackage{hyperref}
\usepackage{times}
\usepackage{amsmath}
\usepackage{graphicx}
\usepackage{subfigure}
\usepackage{placeins}
\usepackage{hyperref}
\usepackage{gensymb}
\usepackage{upgreek}
\usepackage{natbib}

\usepackage{graphicx}
\usepackage{subfigure}
\usepackage{multirow}
\usepackage{comment}
\usepackage{natbib}
\usepackage{hyperref}
\usepackage{mathtools}
\usepackage{mathrsfs}
\usepackage{fontenc}
\usepackage{color}
\usepackage{url}
\usepackage{hyperref}
\usepackage{gensymb}
\usepackage{pifont}
\graphicspath{{./}}

\newcommand       \K            {\,{\rm K}}
\newcommand       \mum          {{\rm \mu m}}

\submitjournal{ApJL}

\shorttitle{Circumnuclear Region of NGC\,7469 Viewed with JWST}
\shortauthors{Zhang \& Ho}

\begin{document}

\title{The Interaction between AGN and Starburst Activity in the Circumnuclear Region of NGC\,7469 as Viewed with JWST}

\author[0000-0003-4937-9077]{Lulu Zhang}
\affiliation{Kavli Institute for Astronomy and Astrophysics, Peking University, Beijing 100871, China}
\affiliation{Department of Astronomy, School of Physics, Peking University, Beijing 100871, China}

\author[0000-0001-6947-5846]{Luis C. Ho}
\affiliation{Kavli Institute for Astronomy and Astrophysics, Peking University, Beijing 100871, China}
\affiliation{Department of Astronomy, School of Physics, Peking University, Beijing 100871, China}

\email{l.l.zhang@pku.edu.cn}

\begin{abstract}
We combine mid-infrared diagnostics obtained from integral-field unit observations taken with MIRI/MRS on JWST with cold molecular gas information derived from ALMA observations of CO(1--0) emission to investigate the star formation rate and efficiency within the central $\sim 1.5\ \rm kpc \times1.3\ \rm kpc$ region of the Seyfert~1 galaxy NGC\,7469 on $\sim 100$~pc scales. The active nucleus leaves a notable imprint on its immediate surroundings by elevating the temperature of the warm molecular gas, driving an ionized gas outflow on sub-kpc scales, and selectively destroying small dust grains. These effects, nevertheless, have relatively little impact on the cold circumnuclear medium or its ability to form stars. Most of the star formation in NGC\,7469 is confined to a clumpy starburst ring, but the star formation efficiency remains quite elevated even for the nuclear region that is most affected by the active nucleus.
\end{abstract}


\section{Introduction}

Grasping the nature of star formation within different environments is fundamental for understanding galaxy formation and evolution. Deciphering star formation activity in galaxies of diverse types and in different stages of their evolution involves consideration of multiple facets of their lifecycle. Of particular recent interest is the potential role played by active galactic nuclei (AGNs) in impacting the star formation properties of galaxies. Widely regarded to coevolve with their galaxies \citep{Kormendy & Ho 2013}, supermassive black holes can influence their hosts and large-scale surroundings in a variety of ways (e.g., \citealt{Fabian 2012, Heckman & Best 2014, Padovani et al. 2017}). Supermassive black holes, while they accrete vigorously and manifest themselves as luminous AGNs and quasars \citep{Rees 1984, Shen & Ho 2014}, release sufficient radiative energy to expel the surrounding gas through ``quasar-mode'' feedback (e.g., \citealt{Di Matteo et al. 2005, Hopkins et al. 2008}). At the opposite extreme, highly sub-Eddington AGNs such as low-ionization nuclear emission-line regions \citep{Heckman 1980, Ho 2008}, which have extremely weak luminosity and very low radiative efficiency \citep{Ho et al. 2003, Ho 2009}, may interact with their environment mainly via ``kinetic-mode'' feedback(e.g., \citealt{McNamara & Nulsen 2007, Weinberger et al. 2017, Dave et al. 2019}), given the propensity for these systems to exhibit increasingly prominent signatures of jet-like emission (e.g., \citealt{Ho 2002, Terashima & Wilson 2003, Sikora et al. 2007}). Although AGN feedback has been incorporated into many state-of-the-art cosmological simulations of galaxy evolution (e.g., \citealt{Schaye et al. 2015, Pillepich et al. 2018, Dave et al. 2020}), debate persists as to whether this mechanism effectively regulates the gas content (e.g., \citealt{Shangguan et al. 2018, Shangguan & Ho 2019, Molina et al. 2022, Zhuang & Ho 2020, Ward et al. 2022}) and star formation activity (e.g., \citealt{Xie et al. 2021}) in galaxies. And if so, which mode operates? AGN feedback can be negative, by expelling gas from galaxies and curtailing their ability or efficiency to form stars (e.g., \citealt{Schawinski et al. 2007, Page et al. 2012, Cheung et al. 2016, Harrison 2017}), or positive, by facilitating star formation through gas compression by outflows (e.g., \citealt{Zubovas et al. 2013, Maiolino et al. 2017, Gallagher et al. 2019, Zhuang et al. 2021}).

(Ultra)luminous infrared galaxies [(U)LIRGs; \citealt{Sanders & Mirabel 1996}] provide an ideal laboratory to test the AGN feedback paradigm. They also serve as excellent local analogs of high-redshift ($z \approx 2$) submillimeter galaxies (e.g., \citealt{Rujopakarn et al. 2011, Bellocchi et al. 2022}). The aftermath of gas-rich, major mergers, (U)LIRGs derive their power mostly from starburst activity, with an admixture of contribution from a dust-enshrouded AGN (e.g., \citealt{Genzel et al. 1998, Shangguan et al. 2019}). In the merger-driven evolutionary scenario of galaxies (e.g., \citealt{Hopkins et al. 2006, Hopkins et al. 2008}), gas funneled to the center of the merger remnant fuels a central starburst and obscured black hole growth, until energy feedback clears enough gas and dust to reveal an optically visible quasar \citep{Sanders et al. 1988a, Sanders et al. 1988b}, and ultimately a quiescent elliptical galaxy \citep{Kormendy & Sanders 1992, Di Matteo et al. 2005}. In view of the complex internal substructure and severe extinction of (U)LIRGs, infrared (IR) observations, especially when spatially resolved, afford the best opportunity to diagnose the physical nature of these complicated environments. As reviewed by \cite{Armus et al. 2020}, previous studies of IR-luminous galaxies, both near and far, have relied heavily on the Spitzer Space Telescope \citep{Werner et al. 2004}. With over one order of magnitude improvement in sensitivity and spatial and spectral resolution compared to Spitzer, the James Webb Space Telescope (JWST; \citealt{Rigby et al. 2023}) presents an unprecedented opportunity to promote our understanding of the evolutionary processes of galaxies. The Medium Resolution Spectrograph (MRS; \citealt{{Wells et al. 2015}}) on the Mid-Infrared Instrument (MIRI; \citealt{Rieke et al. 2015}) is especially promising, given its sensitivity of $\sim 0.1-10$ mJy, full-width at half maximum (FWHM) angular resolution $\sim 0\farcs30-1\farcs10$, and spectral resolution $R \approx 4000-1500$ over the wavelength range $\sim 5-28\, \mum$ (\citealt{Glasse et al. 2015, Labiano et al. 2021}).

At a luminosity distance\footnote{We adopt a redshift of 0.01627 \citep{Springob et al. 2005} and a concordance cosmology with $H_0=70$~km~s$^{-1}$~Mpc$^{-1}$, $\Omega_m=0.3$, and $\Omega_{\Lambda}=0.7$ \citep{Hinshaw et al. 2013}.} of 70.6~Mpc, the SBa spiral galaxy NGC\,7469, one of the six sources originally observed by \cite{Seyfert 1943}, hosts a type~1 (broad-line) AGN with a black hole mass of $\sim 1 \times 10^7\,M_\odot$ estimated from reverberation mapping of the H$\beta$ line \citep{Peterson et al. 2014, Lu et al. 2021} and an X-ray (2--10~keV) luminosity of $5\times 10^{43}\, {\rm erg\ s^{-1}}$ \citep{Ricci et al. 2017}. Included in the Great Observatories All-Sky LIRG Survey (GOALS), NGC\,7469 has an IR luminosity of $10^{11.65}\ L_{\odot}$ \citep{Armus et al. 2009} powered by both its AGN (\citealt{Inami et al. 2013, Stierwalt et al. 2014}) and substantial additional star formation activity ($\sim 40-80\,M_\odot ~\rm yr^{-1}$; \citealt{Perez-Torres et al. 2009, Pereira-Santaella et al. 2011, Mehdipour et al. 2018}), which is concentrated in a nuclear region with radius $\sim 60$~pc \citep{Davies et al. 2007} and a circumnuclear starburst ring of radius $\sim 500$~pc (\citealt{Wilson et al. 1991, Miles et al. 1994, Diaz-Santos et al. 2007, Song et al. 2021}). VLBI observations at 18~cm resolve the nuclear radio emission into several compact sources aligned on an east-west direction of linear extent $\sim 60$~pc \citep{Lonsdale et al. 2003}, which seem to emanate from a disk-like dust structure resolved at 12.5\,$\mum$ \citep{Soifer et al. 2003}. After properly excluding the host galaxy contribution to the optical light, NGC\,7469 qualifies as a radio-loud AGN \citep{Ho & Peng 2001} by the criterion of \cite{Kellermann et al. 1989}. At infrared and millimeter wavelengths, complex ridges connect the nucleus and the circumnuclear ring \citep{Genzel et al. 1995, Davies et al. 2004, Izumi et al. 2020}, including a possible nuclear bar-like structure \citep{Mazzarella et al. 1994}. Optical integral-field spectroscopy also show ionized outflows on kpc scales, possibly driven by the coupled effects of star formation and AGN feedback \citep{Muller-Sanchez et al. 2011, Robleto-Orus et al. 2021, Xu & Wang 2022}. 

The primary goal of this study is to ascertain whether and the manner in which AGN feedback influences the circumnuclear star formation of this well-studied, nearby LIRG, in light of the powerful type~1 AGN it hosts. We take advantage of a recently available set of JWST observations taken with MIRI/MRS to investigate the central $\sim 1.5\ \rm kpc \times 1.3\ \rm kpc$ region of NGC\,7469 on a physical scale of $\sim 100$~pc. The new observations significantly extend previous mid-IR studies based on ground-based facilities (e.g., \citealt{Alonso-Herrero et al. 2014, Esquej et al. 2014, Jensen et al. 2017}).

\begin{figure}[t]
\center{\includegraphics[width=1\linewidth]{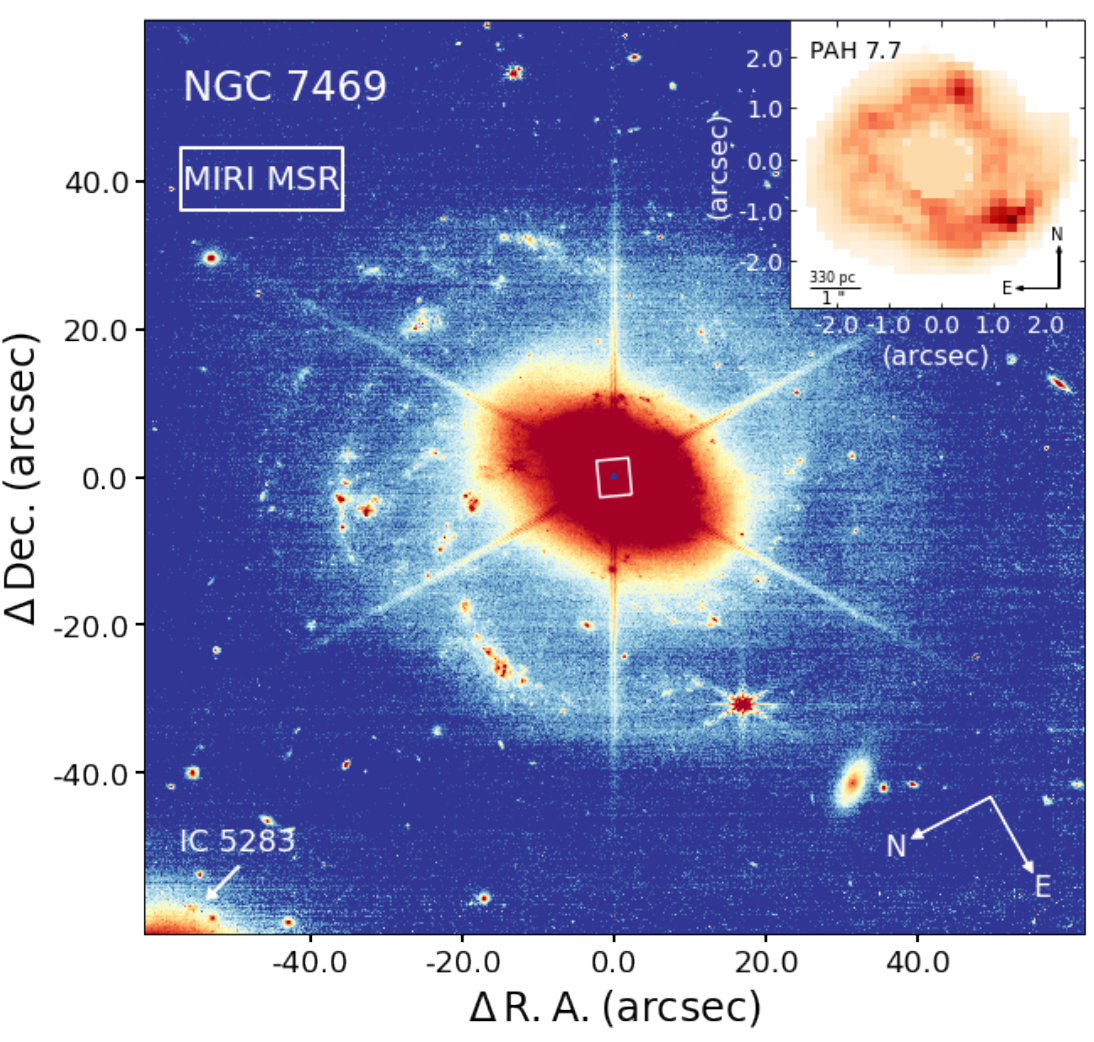}}
\caption{Illustration of the MIRI/MRS coverage (white rectangle) of NGC\,7469, with the NIRCam F335M image as background. The inset shows the distribution of 7.7~$\mum$ PAH emission to illustrate the circumnuclear starburst ring. The bright blob in the lower-left corner is part of the companion galaxy IC\,5283.}\label{fig:coverage}
\end{figure}

\section{Observations and Data Analysis}\label{section:sec2}

NGC\,7469 was included in the GOALS-JWST (ID 1328; co-PIs: L. Armus and A. Evans) Early Release Science program \citep{Lai et al. 2022, U et al. 2022, Armus et al. 2023}. The MIRI/MRS observations were carried out using a set of four integral-field units (channels 1 to 4) that span $4.9-28.1\, \mum$ with three grating settings (short, medium, and long sub-bands). The science exposure time was 444~s, using 40 groups per integration for each sub-band. A four-point dither pattern covers the extended star-forming ring (Figure~\ref{fig:coverage}), and additional dedicated background observations were obtained with the same observational configurations. The MAST Portal\footnote{\url{https://mast.stsci.edu/portal/Mashup/Clients/Mast/Portal.html}} provides four-channel 3D data cubes produced by the standard three-stage JWST pipeline, which consists of the {\tt Detector1}, {\tt Spec2}, and {\tt Spec3} processes. The {\tt Detector1} process applies detector-level corrections with the raw, non-destructive read ramps as input and the uncalibrated slope images (rate files) as output. The {\tt Spec2} step then processes the rate files for other 2D detector-level calibrations, such as WCS assignment, flat-field correction, fringe removal, and wavelength and flux calibration. Finally, {\tt Spec3} processes the resulting calibrated slope images to combined 3D spectral data after background subtraction and cube creation. However, residual fringes remain after standard fringe removal \citep{Argyriou et al. 2020}: these can strongly contaminate the spectral features, especially in the nuclear region. To correct the residual fringing, we begin with the MRS level~2 calibrated data (calibrated slope images), perform master background subtraction following the standard {\tt Spec3} pipeline (\citealt{Bushouse et al. 2022}), and lastly apply an extra step\footnote{\url{https://jwst-pipeline.readthedocs.io/en/latest/jwst/residual_ fringe/main.html}} not implemented in the standard JWST pipeline to correct the low-frequency fringe residuals. The final four-channel 3D data cubes used in this paper are based on the 2D spectral images with the fringe residuals corrected.

\cite{Zhang et al. 2021, Zhang et al. 2022} developed an optimal strategy to extract spatially resolved diagnostic features from mapping-mode observations taken with the Infrared Spectrograph (IRS; \citealt{Houck et al. 2004}) on Spitzer. We adopt the same methodology to analyze the JWST data, showcasing the extraordinary capability of MIRI/MRS observations in studying the interaction between AGN and star formation activity in relatively nearby, well-resolved galaxy environments.

To extract spatially resolved diagnostics on a common physical scale, a key initial step in to ensure that all the slices within the MIRI/MRS data cubes have the same angular resolution. To this end, we begin by convolving all the slices within the first three (1, 2, and 3) channels of the 3D data cubes into a common angular resolution of $\rm FWHM = 0\farcs7$, which is the largest FWHM of the slices within the three channels (Appendix~\ref{sec:AppA}). We focus on the first three channels ($\lambda\approx5-18\, \mum$), which have the highest angular resolution while at the same time cover all the necessary emission features used in this study. Based on the measured FWHMs, we use the {\tt Gaussian2D} function in {\tt astropy.modeling.models} (\citealt{Astropy Collaboration et al. 2013}) to mimic the point-spread function (PSF) of each slice, and we apply the {\tt create\_matching\_kernel} function in {\tt photutils} (\citealt{Bradley et al. 2019}) to construct the convolution kernels. The convolution was performed using the {\tt convolve\_fft} function in {\tt astropy.convolution}.

After convolution, we reproject all the data cubes into the same coordinate frame with a pixel size of 0\farcs35, half of the angular resolution. This allows us to extract 120 spatially resolved spectra from the central $4\farcs55\times3\farcs85$ region of NGC\,7469, in bins of $2 \times 2$ pixels ($0\farcs7 \times 0\farcs7$). Note that each bin share a $0\farcs35 \times 0\farcs7$ region with its most adjacent bins, and the pixel values of the colored maps as shown later are obtained by averaging the measurements of the spatially resolved $0\farcs7 \times 0\farcs7$ bins that cover the corresponding pixel unless specifically noted. Additionally, to mitigate against possible systematics in the convolution process due to our simplified treatment of the PSF, the extracted spectra of the first two channels, which have higher initial resolutions and hence would suffer more from this potential source of systematics, are stitched to the extracted spectra of channel 3. Specifically, each channel 2 spectrum is first stitched to the corresponding channel 3 spectrum based on the overlapping portion of the two, and then the same procedure is repeated for each channel 1 spectrum with respect to the stitched channel 2 spectrum. The mean and standard deviation of the scaling factor is $0.989\pm0.056$ for channel 1 and $0.988\pm0.035$ for channel 2, which is taken into consideration in the final uncertainty of the flux measurements. 

\begin{figure}[t]
\center{\includegraphics[width=1\linewidth]{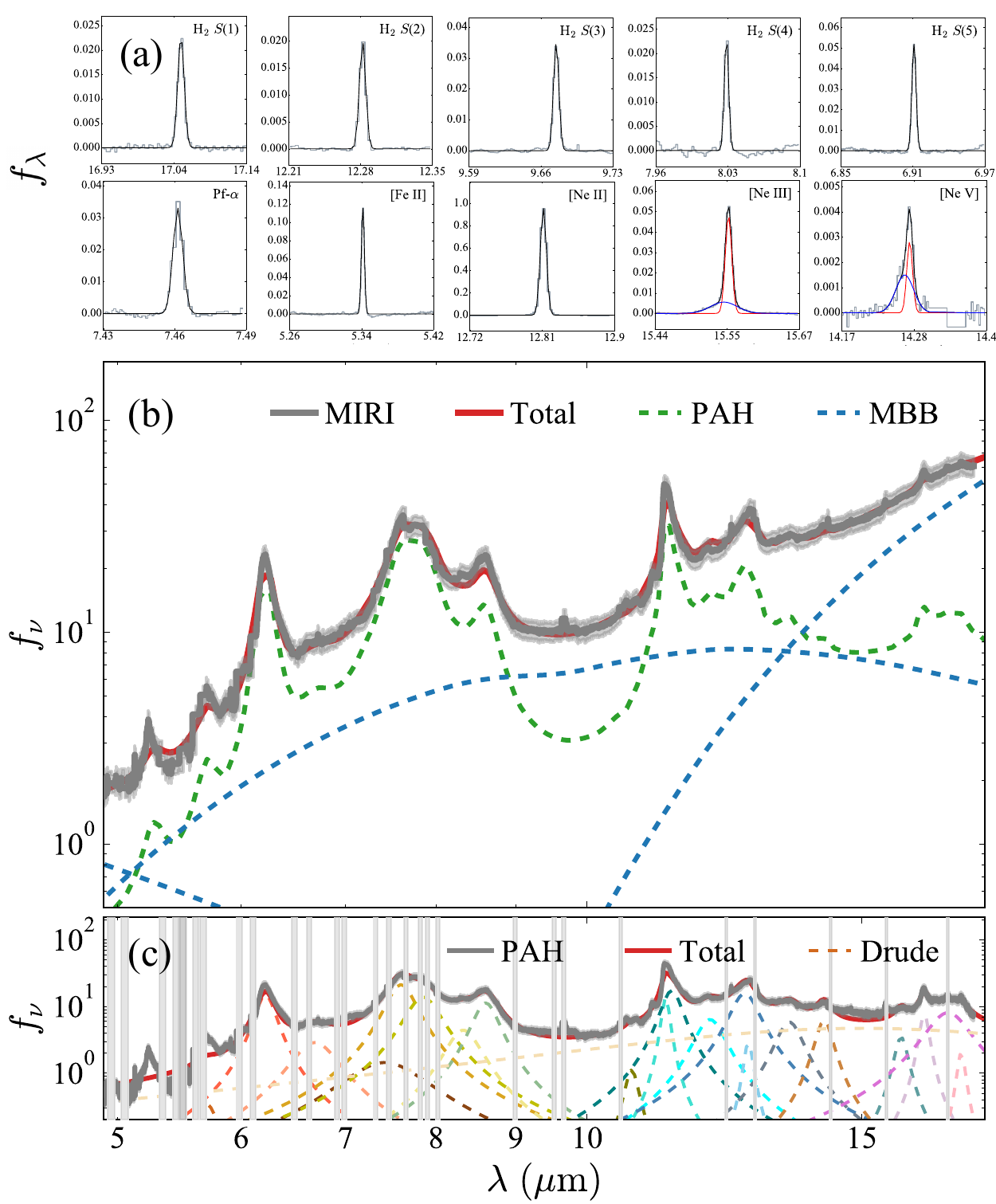}}
\caption{{Panel (a): Measurement of individual emission lines (gray histograms) with either a single (black lines) or a double (blue and red lines) Gaussian function, after subtracting a best-fit local, linear continuum. Panel (b): PAH decomposition, after masking the emission lines, of the MIRI/MRS spectrum (gray solid line) with a multi-component model (red solid line) consisting of a PAH template (green dashed line) and three dust continuum components (blue dashed lines), each represented by a modified blackbody (MBB), corrected for extinction. Panel (c): PAH measurement from the residual PAH spectrum (gray solid line) using a series of Drude profiles (dashed curves) for the individual PAH features, which combine to produce the total model (red solid curve). The residual PAH spectrum was formed from subtracting the best-fit continuum components from the MRS spectrum. The grey stripes mark the regions masked because of emission lines.}\label{fig:sedfit}}
\end{figure}

The MIRI/MRS spectra contain abundant ionic fine-structure lines and molecular hydrogen rotational lines, whose line widths are well resolved under the spectral resolution of these observations. The relative strengths and broadening of these mid-IR lines provide valuable diagnostics of the physical conditions of the gas. This paper focuses on the six hydrogen lines [Pf$\alpha$ and H$_{2}$(0-0)~$S$(1) to H$_{2}$(0-0)~$S$(5)] and four ionic lines ([Fe~{\small II}]~5.340\,$\mum$, [Ne~{\small II}]~12.814\,$\mum$, [Ne~{\small III}]~15.555\,$\mum$, and [Ne~{\small V}]~14.322\,$\mum$) shown in Figure~\ref{fig:sedfit}a and summarized in Tables~\ref{tab:TableLines}--\ref{tab:TableVel}. Most of the lines can be well fit with a single Gaussian function plus a local, linear continuum, except for the high-ionization lines of [Ne~{\small III}] and [Ne~{\small V}], which require two Gaussians. We use the IR extinction curve of \cite{Smith et al. 2007}, as explained below, for extinction correction of the final flux measurement of all the emission lines.

Apart from the above mid-IR emission lines, the broad features from polycyclic aromatic hydrocarbons (PAHs; see reviews by \citealt{Tielens 2008} and \citealt{Li 2020}) provide valuable diagnostics of galaxy properties, such as SFR (\citealt{Shipley et al. 2016, Maragkoudakis et al. 2018, Xie & Ho 2019, Whitcomb et al. 2023, Zhang & Ho 2023b}), molecular gas content (\citealt{Cortzen et al. 2019, Gao et al. 2019, Chown et al. 2021, Leroy et al. 2023, Zhang & Ho 2023a}), and AGN activity (\citealt{Smith et al. 2007, ODowd et al. 2009, Diamond-Stanic & Rieke 2010, Sales et al. 2010, Zhang et al. 2022}). We measure the strengths of the PAH features using the template-fitting method of \cite{Xie et al. 2018}, which was originally devised to treat Spitzer IRS spectra and then further extended by \cite{Zhang et al. 2021, Zhang et al. 2022} to enable spatially resolved analysis of mapping-mode IRS observations. The template-fitting technique can be applied to MIRI/MRS spectra, as illustrated in Figure~\ref{fig:sedfit}b. After masking out the prominent narrow emission lines, the spectrum is fit with a multi-component model consisting of a theoretical PAH template and three modified blackbodies of different temperatures to represent the continuum. All components are subject to attenuation by foreground dust extinction using the IR extinction curve proposed by \citet[][their Equation~4]{Smith et al. 2007}, which is described by a power law plus silicate features peaking at 9.7 and 18\,$\mum$. The extinction curve $\tau(\lambda)$ is scaled by $\tau_{9.7}$, the total extinction at 9.7\,$\mum$ (see Table~\ref{tab:TablePAH}). An additional stellar component for the starlight continuum can also be included in the fit if the near-IR region is well constrained (e.g., \citealt{Zhang & Ho 2023b}), but we omit it here because the contribution of evolved stars to the mid-IR is expected to be small for (U)LIRGs \citep{Paspaliaris et al. 2021}. We measure the integrated PAH luminosity, $L_{\rm PAH\,5-20}$, as the integral of the best-fit PAH template over the region $5-20\,\mum$. The strengths of the most prominent individual PAH features (at 6.2\,$\mum$, 7.7\,$\mum$, and 11.3\,$\mum$; Table~\ref{tab:TablePAH}) are measured from the continuum-subtracted (Figure~\ref{fig:sedfit}c), residual PAH spectrum using a series of Drude profiles \citep{Draine & Li 2007}. The fits are carried out with the Bayesian Markov Chain Monte Carlo procedure {\tt emcee} (see \citealt{Shangguan et al. 2018} for details), with the median and standard deviation of the posterior distribution of each best-fit parameter taken as the final estimate and its corresponding measurement uncertainty. 

\section{Spatially Resolved Analysis}\label{section:sec4}

\subsection{Star Formation Activity}\label{section:sec4.1}

\begin{figure*}[t]
\center{\includegraphics[width=1\linewidth]{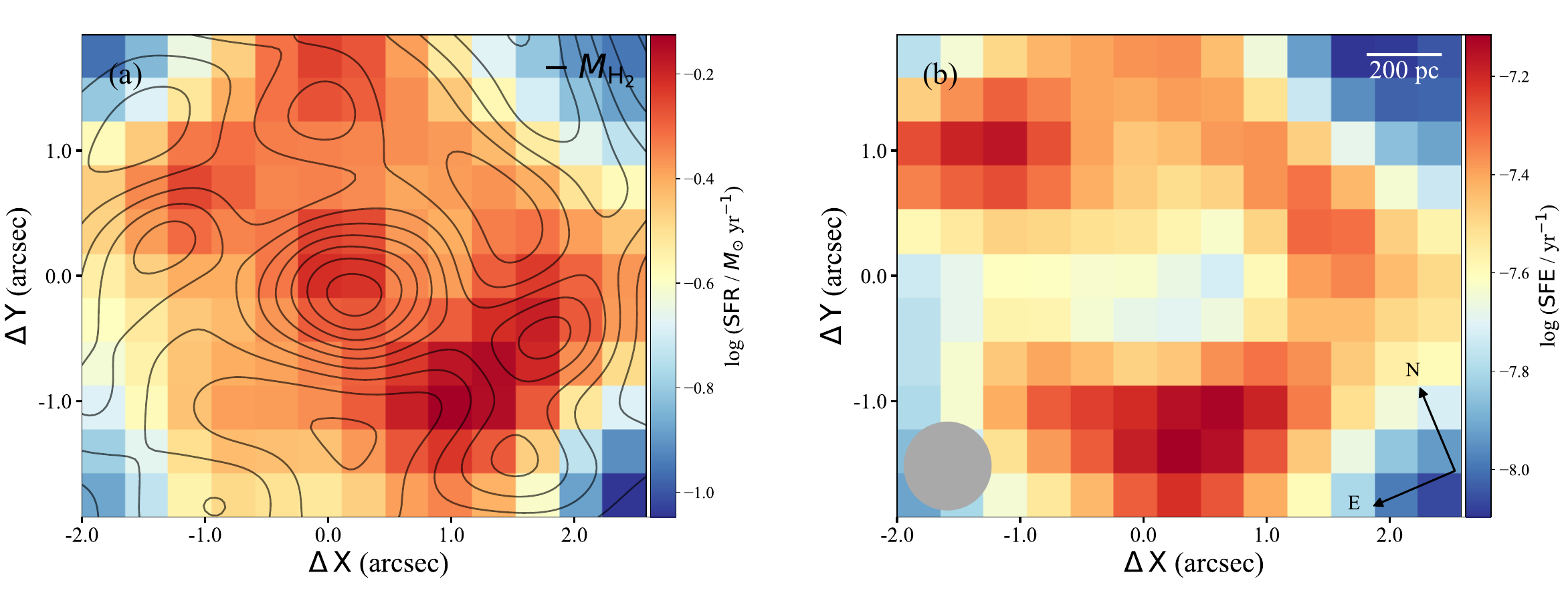}}
\caption{The distribution of (a) SFR and (b) SFE for the central region of NGC\,7469. The black contours in panel (a) indicate the distribution of H$_2$ content derived from the CO(1--0) emission line. The SFE is calculated based on the SFR and H$_2$ content in panel (a). Panel (b) features a filled gray circle to indicate the angular resolution, a compass (with a rotation angle of $23^\circ$), and a scale bar; they are the same for the SFR and H$_2$ maps in panel (a), and for all subsequent colored maps, unless specifically noted.}\label{fig:SFR_SFE}
\end{figure*}

Figure~\ref{fig:SFR_SFE} summarizes the star formation properties of the circumnuclear region of NGC\,7469. We calculate the SFR using the mid-IR fine-structure lines of neon following the prescription of \cite{Zhuang et al. 2019}, who modified the empirical SFR estimator of \cite{Ho & Keto 2007} for star-forming environments by explicitly accounting for the contribution of the AGN to the low-ionization lines. Their photoionization models demonstrate that for a wide range of empirically motivated ionizing spectral energy distributions, the narrow-line region, under conditions of high ionization parameter similar to those of Seyfert galaxies (e.g., \citealt{Ho et al. 1993}), produces a restricted range of [Ne~{\small II}]$+$[Ne~{\small III}]/[Ne~{\small V}] ratio. Under these circumstances, the observed strength of [Ne~{\small V}], if detected as listed in Table~\ref{tab:TableLines}, can be used to remove the contribution of the AGN to [Ne~{\small II}]$+$[Ne~{\small III}], and the SFR can be estimated given a neon abundance.

Assuming that the central region of NGC\,7469 has solar abundances (e.g., \citealt{Alonso-Herrero et al. 2006, Cazzoli et al. 2020}), over the central $\sim 1.5\ \rm kpc \times1.3\ \rm kpc$ region mapped by MRS we obtain an integrated ${\rm SFR} = 55.9\pm 25.7 \,M_\odot ~\rm yr^{-1}$, where the error bars account for a realistic estimate of the systematic uncertainty of the neon-based method ($\sim 0.2$~dex; \citealt{Zhuang et al. 2019}). It is difficult to directly compare our measurement with those of previous studies because of differences in apertures and areas mapped. Nevertheless, our measurement is qualitatively consistent with the ${\rm SFR} = 21\pm2.1 \,M_\odot ~\rm yr^{-1}$ estimated for the circumnuclear ring ($r = 208-963\,{\rm pc}$) from radio continuum emission \citep{Song et al. 2021} and ${\rm SFR} = 10-30 \,M_\odot ~\rm yr^{-1}$ measured over 11 132\,pc-radius circular apertures using the hydrogen recombination line Pf$\alpha$~7.46\,$\mum$ \citep{Lai et al. 2022}. The distribution of SFR shows a central peak, which accounts for only $\sim 1\%$ of the total SFR captured in the mapped region, and an inhomogeneous ring-like structure with several hotspots that are concentrated largely toward the northeast and southwest directions (Figure~\ref{fig:SFR_SFE}a), reminiscent of the morphology seen in the radio (6~cm; \citealt{Wilson et al. 1991}) and mid-IR (11.7\,$\mum$; \citealt{Miles et al. 1994}) bands. A map of Pf$\alpha$ (not shown here), which reflects, at least in part, the distribution of star formation, exhibits a similar structure. 

We overlay on the SFR map the distribution of cold molecular hydrogen gas mass derived from the CO(1--0) observations of ALMA Project 2017.1.00078.S (PI: T. Izumi; \citealt{Izumi et al. 2020}). After extracting the moment 0, 1, and 2 maps from the data cube retrieved from the ALMA archive{\footnote{\url{https://almascience.nao.ac.jp}}} using the {\tt Common Astronomy Software Applications} \citep{CASA Team et al. 2022}, which have an initial resolution of $0\farcs250 \times 0\farcs195$, we convolve moment 0 (CO intensity) map to the same angular resolution as the processed MIRI data cubes (${\rm FWHM} = 0\farcs7$). Following \cite{Izumi et al. 2020}, we adopt a canonical CO-to-H$_2$ conversion factor of ${\rm \alpha_{CO}} = 3.2\ M_{\odot}\ {\rm (K\ km\ s^{-1}\ pc^{2})^{-1}}$ to calculate $M_{\rm H_2}$; this value should be multiplied by a factor of 1.36 to include helium to obtain the total molecular gas \citep{Bolatto et al. 2013}, but throughout we adopt the convention of only considering molecular hydrogen. Over the mapped region we obtain an integrated $M_{\rm H_2} = 1.87\times10^{9} \,M_\odot$. The distribution of H$_2$ gas roughly traces the SFR map. Nevertheless, perhaps not unexpected on these physical scales (e.g., \citealt{Feldmann et al. 2011, Kruijssen & Longmore 2014, Pessa et al. 2021}), the molecular gas clumps along the circumnuclear ring are not spatially coincident with the star-forming hotspots. Intriguingly, molecular gas is piled up toward the nuclear region (within the central $\sim 1\arcsec$), in contrast to the distribution of SFR, which is most prominent in the hotspots on the ring.

\begin{figure}[t]
\center{\includegraphics[width=1\linewidth]{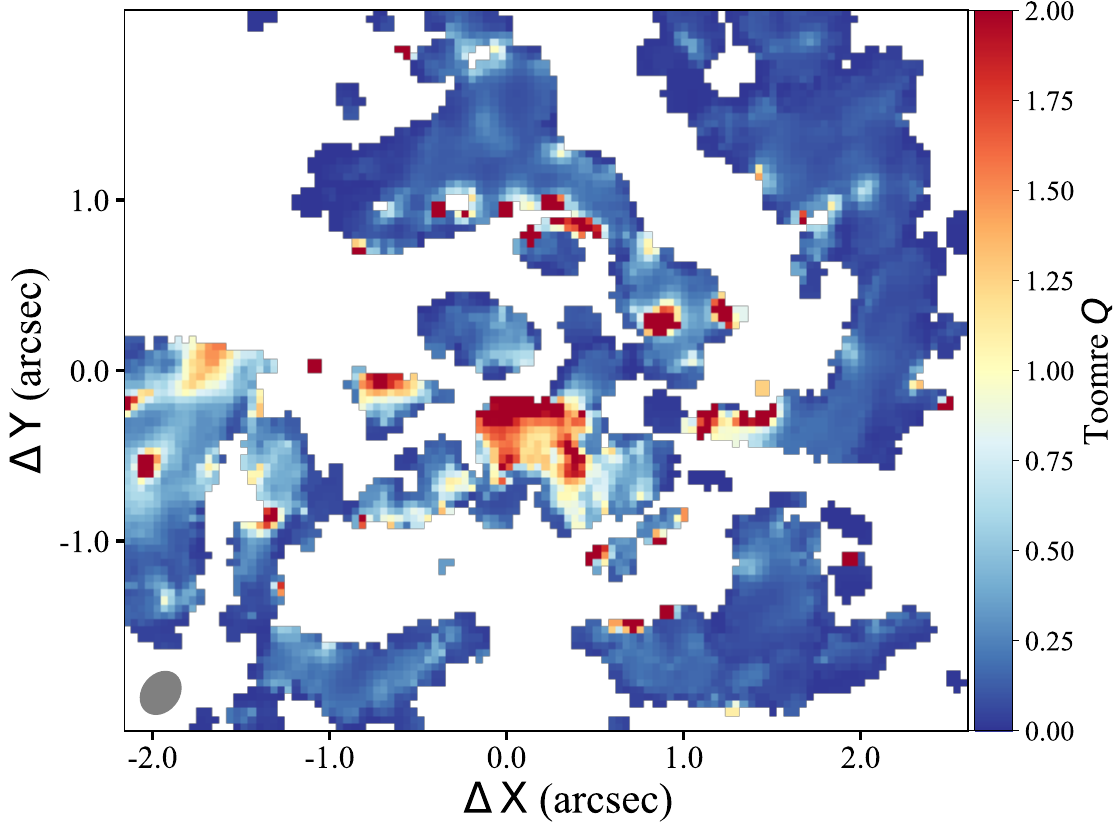}}
\caption{{The distribution of the Toomre $Q$ parameter for the central region of NGC\,7469, calculated from the CO observations. The filled gray ellipse in the bottom-left corner indicates the synthesized beam of the CO map. Note that the distribution is sparse because the velocity and velocity dispersion of the CO line are available only for the pixels with sufficiently strong CO emission.}\label{fig:Toomre_Q}} 
\end{figure}

The mismatch between the gas and young stars is best seen in the distribution of the star formation efficiency, ${\rm SFE} = {\rm SFR}/M_{\rm H_2}$, which presents a minimum at the position of the nucleus, as well as in a gap at $\Delta \rm Y = -0.3$ in the eastern portion of the ring (Figure~\ref{fig:SFR_SFE}b), while the sites of most efficient conversion of gas to stars occur in two concentrations on the circumnuclear ring. The two sites of efficient star formation approximately coincide with the ends of the two spiraling gas arms visible in CO \citep{Izumi et al. 2020}. As observed in other galaxies with prominent bar structures (e.g., \citealt{Kenney et al. 1992, Hsieh et al. 2011}), the bar potential of NGC\,7469 likely facilitates the inner transport of gas, which accumulates in a circumnuclear ring with notable pile-ups in two opposite locations on the ring. The entire circumnuclear region of NGC\,7469 is currently forming stars robustly, with an efficiency that qualifies it as a starburst. The average gas depletion time along the ring is $t_{\rm dep} \equiv {\rm SFE}^{-1} \approx 30 \ {\rm Myr}$, with values as low as $\sim 13\ {\rm Myr}$ in the hotspots. Even the nucleus and the eastern gap region have gas depletion times of $\sim 45\ {\rm Myr}$, which are still significantly shorter than that of star-forming main sequence galaxies ($t_{\rm dep} = 2.35\ {\rm Gyr}$; \citealt{Bigiel et al. 2011}).

\subsection{Stability of the Molecular Gas} \label{section:sec4.2}
To evaluate the stability of the gas, we evaluate \citeauthor{Toomre 1964}'s (1964) parameter $Q = \frac{\sigma \kappa}{\pi G \Sigma_{\rm gas}}$ based on the original CO maps prior to convolution (beam size $0\farcs250\times0\farcs195$; pixel scale $0\farcs04$), with $G$ the gravitational constant, $\Sigma_{\rm gas}$ the mass surface density of H$_2$, $\sigma$ the gas velocity dispersion derived from the CO(1--0) emission line, and the epicyclic frequency $\kappa = 1.41\frac{v(r)}{r}\sqrt{1+\beta}$, with $\beta = d\,{\rm log}\,v(r)/d\,{\rm log}\,r$ (\citealt{Leroy et al. 2008}). For simplicity, we assume a flat rotation curve ($\beta = 0$). The gas velocity at radius $r$, $v(r)$, is calculated from the line-of-sight velocity of the CO(1--0) line according to $v_{\rm los}(x,y) = v_{\rm sys} \ + \ {\rm sin}\ i\ [v_{\rm rot}(r)\ {\rm cos}\ \theta \ +\  v_{\rm exp}(r)\ {\rm sin}\ \theta]$, assuming expansion velocity $v_{\rm exp}(r) = 0$. The azimuthal angle ${\rm cos}\ \theta = \frac{-(x-x_{c})\ {\rm sin}\ \phi\ +\ (y-y_{c})\ {\rm cos}\ \phi}{r}$. Pixels with ${\rm cos}\ \theta < 0.1$ are masked to avoid unphysical values of the approximated $v_{\rm rot}$, and we set the inclination angle $i = 45^{\circ}$ and the position angle $\phi = 308^{\circ}$ \citep{Nguyen et al. 2021}. We do not consider the beam-smearing effect on the velocity dispersion field in light of the small beam size and pixel scale of the CO maps. Beam-smearing effects are usually considered for data cubes of worse resolution (e.g., $\gtrsim 1$~kpc; \citealt{Di Teodoro & Fraternali 2015, Leung et al. 2018, Levy et al. 2018}). Figure~\ref{fig:Toomre_Q} illustrates that, except for a few pixels to the southeast of the nucleus, which roughly correspond to the nuclear minimum SFE region at $(\Delta \rm X, \Delta \rm Y) = (0.3, -0.3)$, $Q$ has values less than 1 throughout most of the central region of NGC\,7469, in agreement with its starburst nature.

\begin{figure}[t]
\center{\includegraphics[width=1\linewidth]{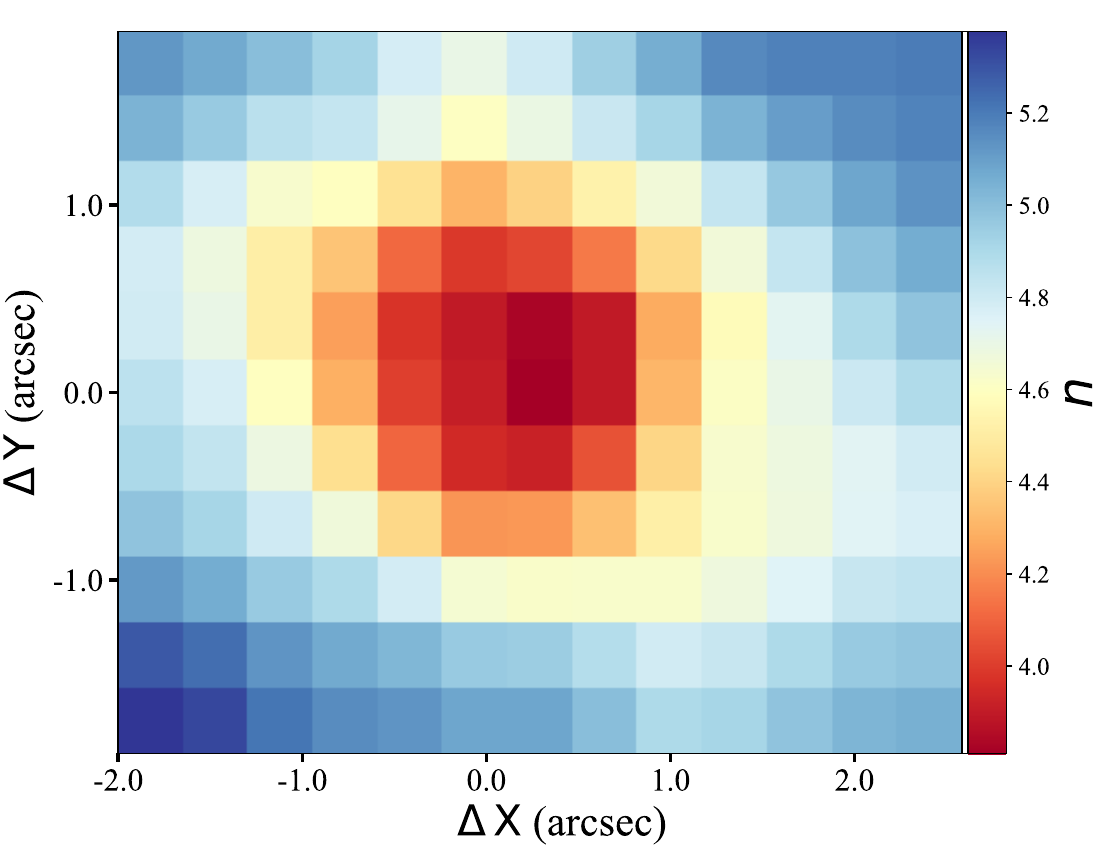}}
\caption{Distribution of the power-law index $n$ for the central region of NGC\,7469, where smaller $n$ corresponds to higher hot gas fraction, with $n$ derived from the H$_2$ rotational lines assuming a power-law distribution of H$_2$ excitation temperatures.}\label{fig:index_n}
\end{figure}

\subsection{Evidence of Gas Heating by the AGN} \label{section:sec4.3}

The relative intensities of the rotational lines of H$_2$ in the electronic ground state provide a powerful diagnostic of the excitation temperature of the warm molecular medium \citep{Rosenthal et al. 2000, Rigopoulou et al. 2002, Roussel et al. 2007}. Assuming a continuous power-law distribution of temperature $T$ and optically thin conditions, the flux of the H$_2$ lines at rotational energy level $J+2$ is related to the column density $N(J) \propto \int_{T_{l}}^{T_{u}}\frac{g(J)}{Z(T)} \, e^{-E(J)/kT}\ T^{-n}dT$, with $g(J)$ the degeneracy value, $E(J)$ the energy of level $J$, $k$ the Boltzman's constant, and $Z(T)$ the partition function at temperature $T$ \citep{Togi & Smith 2016, Appleton et al. 2017}. The power-law index $n$ parameterizes the distribution of gas temperature from $T_l$ to $T_u$, such that smaller values of $n$ value correspond to larger hot gas fraction. Following \cite{Togi & Smith 2016}, we set $T_{u} = 2000$~K and fit for $T_{l}$, which is found to range from $\sim 150$ to 300~K. The distribution of $n$ obtained for the central region of NGC~7469 (Figure~\ref{fig:index_n}) agrees well within the values of $n \approx 4-5$ derived by \cite{Pereira-Santaella et al. 2014} for six local IR-bright Seyfert galaxies, and it falls within the range of $n \approx 2.5-5$ derived for ULIRGs \citep{Zakamska 2010}. Meanwhile, the values of $n$ in NGC~7469 are on average smaller---indicating higher hot gas fraction---than the mean of $n = 4.84\pm0.16$ measured by \cite{Togi & Smith 2016} for nearby star-forming galaxies and low-luminosity AGNs. Most notably, values of $n$ in the circumnuclear region of NGC~7469 gradually but systematically drops toward the nucleus, indicating a central rise in molecular gas temperature. By contrast, the circumnuclear ring is characterized by uniformly lower temperatures. These results suggest that the active nucleus drives some mechanism of gas heating in the central region, which may be the culprit for the depressed SFE in the center.

\begin{figure*}[t]
\center{\includegraphics[width=1\linewidth]{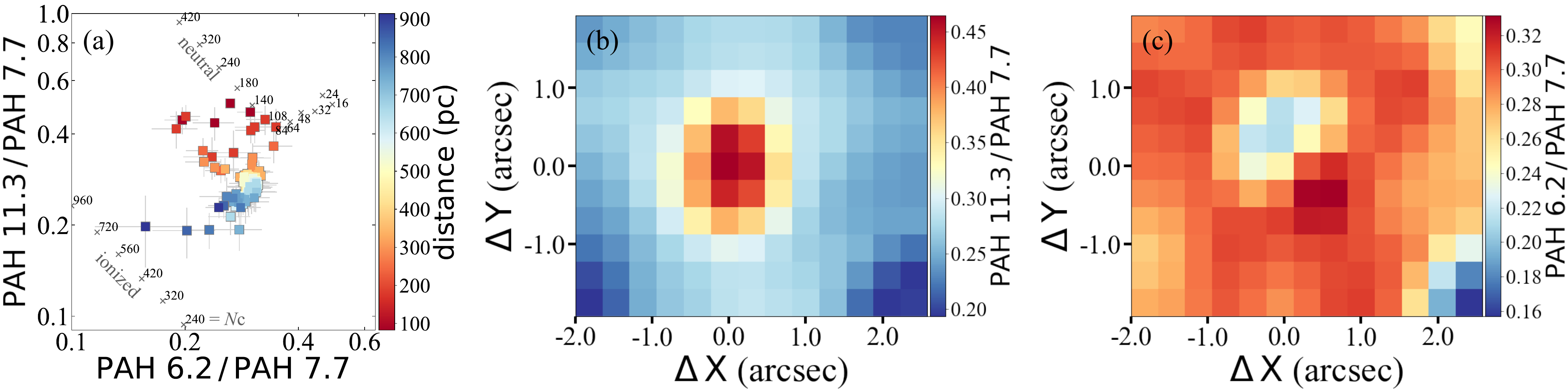}}
\caption{{(a) Diagnostic diagram of PAH band ratios 6.2~$\mum$/7.7~$\mum$ versus 11.3~$\mum$/7.7~$\mum$ for spatially resolved bins of $2 \times 2$ pixels, color-coded according to their locations from the central SFR peak [$(\Delta \rm X, \Delta \rm Y) = (0, 0)$]. The spatial distribution of (b) PAH~11.3\,$\mum$/7.7\,$\mum$ and (c) PAH~6.2\,$\mum$/7.7\,$\mum$.}\label{fig:pahr}}
\end{figure*}

\begin{figure}[b]
\center{\includegraphics[width=1\linewidth]{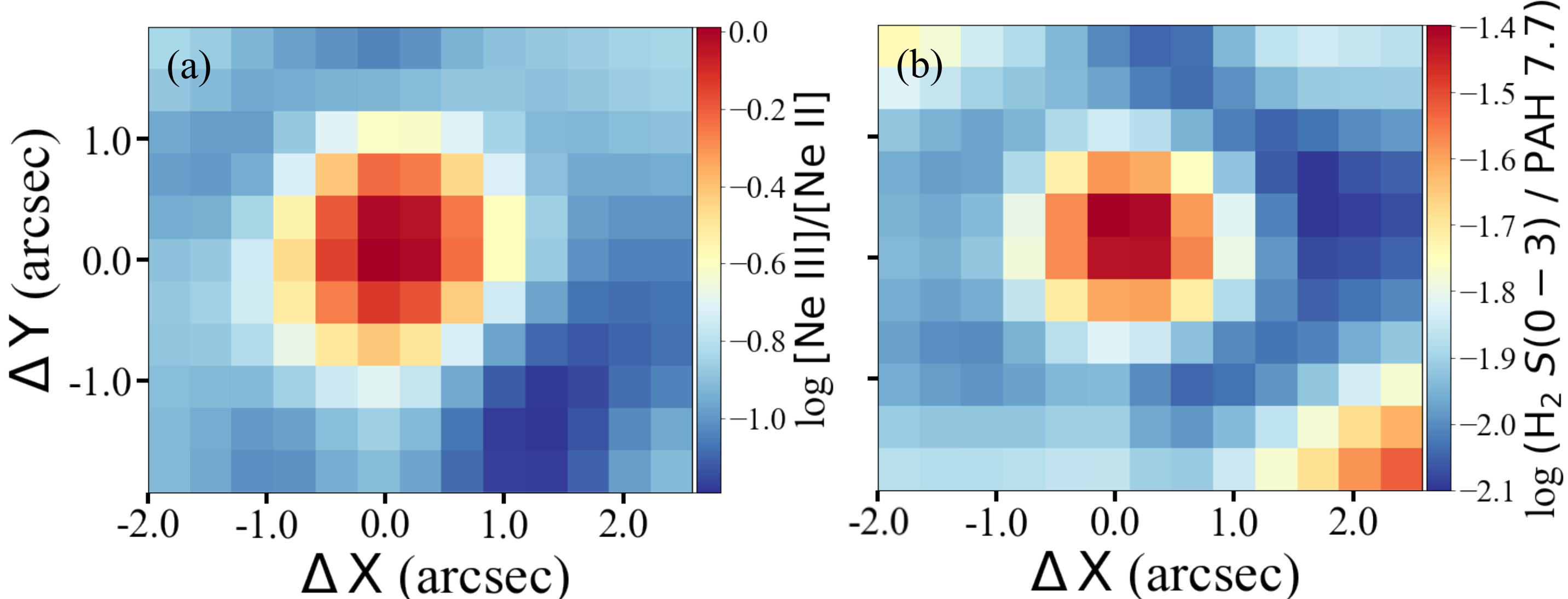}}
\caption{{The spatial distribution of mid-IR line ratio diagnostics (a) [Ne~{\small III}]/[Ne~{\small II}] and (b) H$_2~S(0-3)$/PAH~7.7\,$\mum$.}\label{fig:diags}}
\end{figure}

To further buttress our contention that AGN activity alters the conditions of the gas in the immediate vicinity of the nucleus, we turn to the detailed spectrum of the PAH features, which in AGNs exhibits on average smaller PAH~6.2\,$\mum$/7.7\,$\mum$ and larger PAH~11.3\,$\mum$/7.7\,$\mum$ band ratios compared to star-forming regions as a consequence of radiative processes and shocks \citep{Zhang et al. 2022}. As shown in Figure~\ref{fig:pahr} (panels~a and b), the resolved bins closest to the nucleus all exhibit PAH~11.3\,$\mum$/7.7\,$\mum$ band ratios larger than 0.3, which are distinctly offset from the locus of points at larger radii. \citet{Garcia-Bernete et al. 2022}, using JWST/MRS to study PAH emission in Seyfert galaxies, including NGC~7469, obtained similar results. Such high values of PAH~11.3\,$\mum$/7.7\,$\mum$ are typical of AGNs \citep{Zhang et al. 2022}. The wide range of PAH~6.2 $\mum$/7.7 $\mum$ band ratios reflects the coupled effects of AGN and starburst activity, with the nuclear bins characterized by smaller values typical of AGNs (panel~c).

Apart from having distinctive PAH band ratios, the nuclear bins have elevated [Ne~{\small III}]/[Ne~{\small II}] (Figure~\ref{fig:diags}a), an indicator of the hardness of the radiation field \citep{Thornley et al. 2000} and H$_2~S(0-3)$/PAH~7.7~$\mum$ (Figure~\ref{fig:diags}b), which is sensitive to shocks \citep{Roussel et al. 2007}. ${\rm H_{2}}~S(0-3)$ is the summation of ${\rm H_{2}}~S(0)$ to ${\rm H_{2}}~S(3)$, which is converted from the summation of ${\rm H_{2}}~S(1)$ to ${\rm H_{2}}~S(5)$ by dividing by a factor of 0.9, following \cite{Lai et al. 2022} based on the results of \cite{Habart et al. 2011}. We note, however, that although the nucleus in general has elevated ${\rm H_{2}}\ S(0-3)/{\rm PAH\ 7.7}$ compared to the circumnuclear ring, nevertheless most of the values do not exceed the theoretical threshold that can be generated by photo-dissociation regions [$\log {\rm H_{2}}\ S(0-3)/{\rm PAH\ 7.7} = -1.4$; \citealt{Guillard et al. 2012, Stierwalt et al. 2014}]. Hence, despite the evidence for outflows, as discussed below, the mid-IR line ratios themselves do not point to a dominant role played by shocks in gas heating in the nuclear region.

For completeness, we note that three bins located outside the circumnuclear ring at $(\Delta \rm X, \Delta \rm Y) = (2.3, -1.8)$ stand out for having exceptionally low values of PAH~11.3\,$\mum$/7.7\,$\mum$ and PAH~6.2\,$\mum$/7.7\,$\mum$ and yet high ${\rm H_{2}}\ S(0-3)/{\rm PAH\ 7.7}$. Such line ratios are indicative of mechanical processes like shocks, but their connection to the AGN is unclear.

\begin{figure*}[t]
\center{\includegraphics[width=1\linewidth]{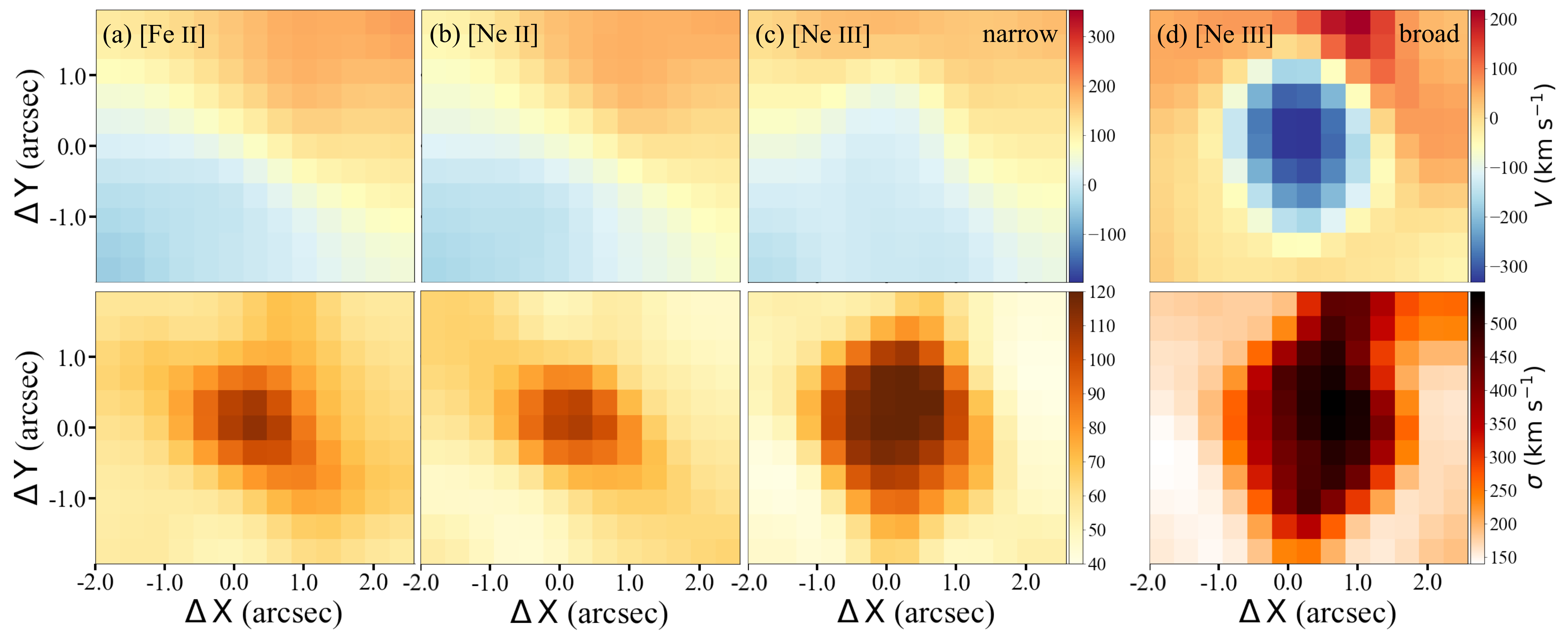}}
\caption{Distribution of line-of-sight velocity (top) and velocity dispersion (bottom) in the central region of NGC\,7469 for the emission lines of (a) [Fe~{\small II}]~5.340\,$\mum$, (b) [Ne~{\small II}]~12.814\,$\mum$, and (c) the narrow and (d) broad component of [Ne~{\small III}]~15.555\,$\mum$. The color bars span the same velocity range in (a)--(c).}
\label{fig:Kin}
\end{figure*}

\subsection{Evidence of an Ionized Gas Outflow from the AGN}\label{section:sec4.4}

In addition to the heating effect deduced from the analysis of the rotational H$_2$ spectrum (Section~\ref{section:sec4.3}), AGN activity can also impact the host galaxy by expelling gas through a wind or outflow. To check whether the central starburst activity of NGC\,7469 is influenced by mechanical processes, we examine the velocity field of the central region of the galaxy using the strongest of the well-isolated ionic emission lines (Figure~\ref{fig:Kin}). The two low-ionization lines of [Fe~{\small II}] and [Ne~{\small II}] have rather regular kinematics, showing a velocity field dominated by large-scale rotation with a projected amplitude of approximately $\pm 120\,\rm km\ s^{-1}$ (panels~a and b). The velocity dispersions, after correcting for instrumental broadening according to the line-spread function of \cite{Labiano et al. 2021}, are centrally peaked and slightly extended along the minor axis of the galaxy (at position angle $\Theta \approx 35^{\circ}$). The region with spatially extended velocity dispersion roughly corresponds to the nuclear bar-like structure of NGC\,7469, which is visible on a similar physical scale in near-IR and CO emission \citep{Mazzarella et al. 1994, Izumi et al. 2020}. 

By contrast, the higher ionization line of [Ne~{\small III}] behaves markedly differently. As previously noted (Figure~\ref{fig:sedfit}), [Ne~{\small III}] can be decomposed into a narrow and a broad component. While the velocity field for the narrow component of [Ne~{\small III}] generally follows that of the low-ionization lines, the velocities near the nuclear region are severely distorted, and the velocity dispersions reach values as large as $\sigma \approx 120\,\rm km\ s^{-1}$ over an extended region of $\sim 2\arcsec$ along $\Theta \approx 115^{\circ}$ (panel~c). Most intriguingly, over roughly the same spatial extend of $\sim 2\arcsec$, we detect in the broad component of [Ne~{\small III}] prominent blueshifts of up to $-300\,\rm km\ s^{-1}$ (panel~d), adjacent to which toward the northwest is a minor component redshifted by $+200\,\rm km\ s^{-1}$. The redshifted component might be the backside of a bipolar outflow, which has lower intensity because of dust extinction. Very large velocity dispersions, in excess of $\sigma \gtrsim 500\,\rm km\ s^{-1}$, appear throughout the central region, but with a preference along $\Theta \approx 115^{\circ}$. These characteristics of the [Ne~{\small III}] line strongly suggest a sub-kpc ionic gas outflow.

\section{Summary}\label{section:sec5.2}

Using integral-field unit observations taken by MIRI/MRS on JWST, we study the interaction between AGN and starburst activity within the central $1.5\,\rm kpc \times1.3\ \rm kpc$ region of the Seyfert~1 galaxy NGC\,7469 on a scale of $\sim 100$~pc. The distribution of SFR derived from the mid-IR fine-structure lines of [Ne~{\small II}] and [Ne~{\small III}] delineate two zones of star formation activity, one centered on the nucleus and the other confined to a circumnuclear ring. We take advantage of ALMA CO(1--0) observations to obtain the distribution of cold molecular hydrogen gas, which, when combined with the SFR, translates into a map of the SFE. The MIRI/MRS bandpass covers powerful diagnostics from the rotational H$_2$ lines and PAH features, and the spectral resolution yields new insights from the kinematics of the ionic lines of low and high ionization. 

The main results can be summarized as follows:

\begin{enumerate}

\item The central region of NGC\,7469 experiences intense starburst activity, which is confined largely to a clumpy circumnuclear ring but also manifests itself, albeit to a lower degree, toward the nucleus.

\item The active nucleus drives an ionized gas outflow on sub-kpc scales, detectable in the broad component of the [Ne~{\small III}] line as emission with blueshifts of up to $-300\,\rm km\ s^{-1}$ and redshifts of $+200\,\rm km\ s^{-1}$, as well as velocity dispersions that reach $\sim 500\,\rm km\ s^{-1}$.

\item The active nucleus significantly influences the properties of the molecular gas in its immediate surroundings, by sharply raising the gas temperature as inferred from the rotational H$_2$ lines and by selectively destroying small dust grains as evidenced by the PAH band ratios. While shocks may be present, radiative heating appears to dominate.

\end{enumerate}

Despite the observable manifestations of NGC\,7469's highly accreting supermassive black hole, AGN feedback overall has a negligible impact on the cold circumnuclear medium or its ability to form stars. A formal analysis of the stability of the molecular gas demonstrates that Toomre's $Q$ parameter is largely less than 1 throughout the circumnuclear region. The circumnuclear ring of the galaxy remains a healthy starburst, and even the very nuclear region itself forms stars with substantial efficiency. These conclusions echo similar findings reported in other studies of nearby AGNs (e.g., \citealt{Xie et al. 2021, Zhuang et al. 2021, Molina et al. 2022, Molina et al. 2023, Zhuang & Ho 2022}).

\newpage

\acknowledgments
We thank the anonymous referee for helpful comments and suggestions. We thank Juan Molina (KIAA) and Masafusa Onoue (KIAA) for organizing the JWST Hackathon to learn about reducing and analyzing data of the first JWST observations. We thank Yuanqi Liu (SHAO) for helping to generate the CO maps from the ALMA data cube. This work was supported by the National Science Foundation of China (11721303, 11991052, 12011540375, 12233001), the National Key R\&D Program of China (2022YFF0503401), and the China Manned Space Project (CMS-CSST-2021-A04, CMS-CSST-2021-A06). The JWST data presented in this paper were obtained from the Mikulski Archive for Space Telescopes (MAST) at the Space Telescope Science Institute. The specific observations analyzed can be accessed via \dataset[https://doi.org/10.17909/18x3-kv08]{https://doi.org/10.17909/18x3-kv08}. This paper makes use of the following ALMA data: ADS/JAO.ALMA\#2017.1.00078.S. ALMA is a partnership of ESO (representing its member states), NSF (USA) and NINS (Japan), together with NRC (Canada), MOST and ASIAA (Taiwan), and KASI (Republic of Korea), in cooperation with the Republic of Chile. The Joint ALMA Observatory is operated by ESO, AUI/NRAO and NAOJ. This research made use of {\tt Montage}. It is funded by the National Science Foundation under Grant Number ACI-1440620, and was previously funded by the National Aeronautics and Space Administration's Earth Science Technology Office, Computation Technologies Project, under Cooperative Agreement Number NCC5-626 between NASA and the California Institute of Technology.

\startlongtable
\setlength{\tabcolsep}{3pt}
\begin{deluxetable*}{ccccccccc}
\tabletypesize{\scriptsize}
\tablecolumns{9}
\tablecaption{Spatially Resolved Measurements of Ionic Emission Lines, SFR, and H$_2$ Mass}
\tablehead{
\colhead{Region} & \colhead{Coordinate} & \colhead{log $f_{\rm Pf\alpha}$} & \colhead{log $f_{\rm [Fe~{\scriptsize II}]}$} & \colhead{log $f_{\rm [Ne~{\scriptsize II}]}$} & \colhead{log $f_{\rm [Ne~{\scriptsize III}]}$} & \colhead{log $f_{\rm [Ne~{\scriptsize V}]}$} & \colhead{log ${\rm SFR}$} & \colhead{log ${M_{\rm H_2}}$} \\
\colhead{} &  \colhead{(pixel)} & \colhead{($\rm erg~s^{-1}~cm^{-2}$)} & \colhead{($\rm erg~s^{-1}~cm^{-2}$)} & \colhead{($\rm erg~s^{-1}~cm^{-2}$)} & \colhead{($\rm erg~s^{-1}~cm^{-2}$)} & \colhead{($\rm erg~s^{-1}~cm^{-2}$)} & \colhead{($M_{\odot}~{\rm yr^{-1}}$)} & \colhead{($M_{\odot}$)} \\
\colhead{(1)} & \colhead{(2)} & \colhead{(3)} & \colhead{(4)} & \colhead{(5)} & \colhead{(6)} & \colhead{(7)} & \colhead{(8)} & \colhead{(9)}}
\startdata
NGC7469\_bin\_1 & ($-$0.5, $-$0.5) & $-$14.836 $\pm$ 0.045 & $-$14.355 $\pm$ 0.039 & $-$13.162 $\pm$ 0.039 & $-$13.203 $\pm$ 0.039 & $-$13.403 $\pm$ 0.039 & 0.330 $\pm$ 0.043 & 8.007 $\pm$ 0.018 \\
NGC7469\_bin\_2 & ($-$0.5, 0.5) & $-$14.697 $\pm$ 0.033 & $-$14.354 $\pm$ 0.031 & $-$13.148 $\pm$ 0.031 & $-$13.133 $\pm$ 0.031 & $-$13.354 $\pm$ 0.032 & 0.359 $\pm$ 0.034 & 7.939 $\pm$ 0.021 \\
NGC7469\_bin\_3 & (0.5, $-$0.5) & $-$14.710 $\pm$ 0.043 & $-$14.274 $\pm$ 0.024 & $-$13.098 $\pm$ 0.024 & $-$13.087 $\pm$ 0.023 & $-$13.301 $\pm$ 0.025 & 0.405 $\pm$ 0.026 & 8.092 $\pm$ 0.015 \\
NGC7469\_bin\_4 & (0.5, 0.5) & $-$14.588 $\pm$ 0.022 & $-$14.268 $\pm$ 0.006 & $-$13.074 $\pm$ 0.006 & $-$13.014 $\pm$ 0.006 & $-$13.260 $\pm$ 0.009 & 0.450 $\pm$ 0.007 & 8.022 $\pm$ 0.017 \\
NGC7469\_bin\_5 & ($-$1.5, $-$0.5) & $-$14.994 $\pm$ 0.057 & $-$14.482 $\pm$ 0.056 & $-$13.296 $\pm$ 0.057 & $-$13.587 $\pm$ 0.056 & $-$13.826 $\pm$ 0.057 & 0.181 $\pm$ 0.054 & 7.855 $\pm$ 0.025 \\
NGC7469\_bin\_6 & ($-$1.5, 0.5) & $-$14.958 $\pm$ 0.050 & $-$14.486 $\pm$ 0.049 & $-$13.274 $\pm$ 0.049 & $-$13.532 $\pm$ 0.049 & $-$13.775 $\pm$ 0.050 & 0.207 $\pm$ 0.047 & 7.815 $\pm$ 0.028 \\
NGC7469\_bin\_7 & ($-$0.5, $-$1.5) & $-$14.943 $\pm$ 0.039 & $-$14.406 $\pm$ 0.037 & $-$13.252 $\pm$ 0.039 & $-$13.503 $\pm$ 0.035 & $-$13.725 $\pm$ 0.036 & 0.225 $\pm$ 0.037 & 7.879 $\pm$ 0.024 \\
NGC7469\_bin\_8 & ($-$0.5, 1.5) & $-$14.843 $\pm$ 0.045 & $-$14.433 $\pm$ 0.043 & $-$13.231 $\pm$ 0.043 & $-$13.324 $\pm$ 0.043 & $-$13.591 $\pm$ 0.043 & 0.275 $\pm$ 0.043 & 7.737 $\pm$ 0.033 \\
NGC7469\_bin\_9 & (0.5, $-$1.5) & $-$14.918 $\pm$ 0.036 & $-$14.348 $\pm$ 0.035 & $-$13.196 $\pm$ 0.037 & $-$13.420 $\pm$ 0.034 & $-$13.636 $\pm$ 0.035 & 0.282 $\pm$ 0.035 & 7.965 $\pm$ 0.019 \\
NGC7469\_bin\_10 & (0.5, 1.5) & $-$14.789 $\pm$ 0.032 & $-$14.379 $\pm$ 0.028 & $-$13.194 $\pm$ 0.029 & $-$13.231 $\pm$ 0.028 & $-$13.517 $\pm$ 0.028 & 0.326 $\pm$ 0.028 & 7.813 $\pm$ 0.028 \\
\enddata
\tablecomments{\footnotesize Col. (1): The resolved $2 \times2$ pixel ($0\farcs7 \times 0\farcs7$) bins, ordered according to their radial distance from the peak of the nuclear SFR (the origin of the coordinates in Figure~3: $\alpha = \rm 23^{h}03^{m}15\fs626$, $\delta = \rm 08\degree52\arcmin26\farcs16$). Col. (2): Offset of the bin center relative to the origin of the coordinates, in units of the pixel scale of 0\farcs35. Cols. (3)--(7): Extinction-corrected flux of the emission lines Pf$\alpha\,7.46\,\mum$, [Fe~{\scriptsize II}]~5.340\,$\mum$, [Ne~{\scriptsize II}]~12.814\,$\mum$, [Ne~{\scriptsize III}]~15.555\,$\mum$, and [Ne~{\scriptsize V}]~14.322\,$\mum$; the uncertainty is the quadrature sum of the measurement uncertainty from MCMC fitting and the systematic uncertainty from the scaling of each spectrum. Col. (8): The SFR based on the three neon lines. Col. (9): The molecular hydrogen mass, derived from CO(1--0) emission assuming a CO-to-H$_2$ conversion factor of ${\rm \alpha_{CO}} = 3.2\ M_{\odot}\ {\rm (K\ km\ s^{-1}\ pc^{2})^{-1}}$. (This table is available in its entirety in machine-readable form.)}
\label{tab:TableLines}
\end{deluxetable*}

\startlongtable
\setlength{\tabcolsep}{5pt}
\begin{deluxetable*}{ccccccc}
\tabletypesize{\scriptsize}
\tablecolumns{7}
\tablecaption{Spatially Resolved Measurements of H$_2$ Rotational Lines and Power-law Index $n$}
\tablehead{
\colhead{Region} & \colhead{log $f_{{\rm H_{2}}\ S(1)}$} & \colhead{log $f_{{\rm H_{2}}\ S(2)}$} & \colhead{log $f_{{\rm H_{2}}\ S(3)}$} & \colhead{log $f_{{\rm H_{2}}\ S(4)}$} & \colhead{log $f_{{\rm H_{2}}\ S(5)}$} & \colhead{$n$} \\
\colhead{} &\colhead{($\rm erg~s^{-1}~cm^{-2}$)} & \colhead{($\rm erg~s^{-1}~cm^{-2}$)} & \colhead{($\rm erg~s^{-1}~cm^{-2}$)} & \colhead{($\rm erg~s^{-1}~cm^{-2}$)} & \colhead{($\rm erg~s^{-1}~cm^{-2}$)} & \colhead{} \\
\colhead{(1)} & \colhead{(2)} & \colhead{(3)} & \colhead{(4)} & \colhead{(5)} & \colhead{(6)} & \colhead{(7)}}
\startdata
NGC7469\_bin\_1 & $-$14.401 $\pm$ 0.048 & $-$14.691 $\pm$ 0.040 & $-$14.423 $\pm$ 0.039 & $-$14.528 $\pm$ 0.039 & $-$14.191 $\pm$ 0.039 & 3.915 $\pm$ 0.776 \\
NGC7469\_bin\_2 & $-$14.302 $\pm$ 0.037 & $-$14.622 $\pm$ 0.033 & $-$14.328 $\pm$ 0.032 & $-$14.431 $\pm$ 0.031 & $-$14.111 $\pm$ 0.031 & 3.942 $\pm$ 0.767 \\
NGC7469\_bin\_3 & $-$14.318 $\pm$ 0.045 & $-$14.670 $\pm$ 0.029 & $-$14.319 $\pm$ 0.025 & $-$14.408 $\pm$ 0.025 & $-$14.076 $\pm$ 0.023 & 3.870 $\pm$ 0.784 \\
NGC7469\_bin\_4 & $-$14.209 $\pm$ 0.025 & $-$14.598 $\pm$ 0.015 & $-$14.216 $\pm$ 0.008 & $-$14.313 $\pm$ 0.008 & $-$14.000 $\pm$ 0.005 & 3.917 $\pm$ 0.782 \\
NGC7469\_bin\_5 & $-$14.701 $\pm$ 0.057 & $-$14.759 $\pm$ 0.056 & $-$14.572 $\pm$ 0.056 & $-$14.773 $\pm$ 0.056 & $-$14.463 $\pm$ 0.056 & 4.130 $\pm$ 0.998 \\
NGC7469\_bin\_6 & $-$14.674 $\pm$ 0.049 & $-$14.713 $\pm$ 0.049 & $-$14.507 $\pm$ 0.049 & $-$14.696 $\pm$ 0.049 & $-$14.387 $\pm$ 0.049 & 4.044 $\pm$ 1.030 \\
NGC7469\_bin\_7 & $-$14.603 $\pm$ 0.078 & $-$14.811 $\pm$ 0.038 & $-$14.585 $\pm$ 0.037 & $-$14.761 $\pm$ 0.037 & $-$14.443 $\pm$ 0.036 & 3.987 $\pm$ 0.705 \\
NGC7469\_bin\_8 & $-$14.437 $\pm$ 0.048 & $-$14.660 $\pm$ 0.043 & $-$14.362 $\pm$ 0.043 & $-$14.521 $\pm$ 0.043 & $-$14.224 $\pm$ 0.043 & 3.815 $\pm$ 0.671 \\
NGC7469\_bin\_9 & $-$14.492 $\pm$ 0.050 & $-$14.793 $\pm$ 0.038 & $-$14.524 $\pm$ 0.035 & $-$14.671 $\pm$ 0.035 & $-$14.354 $\pm$ 0.034 & 4.013 $\pm$ 0.704 \\
NGC7469\_bin\_10 & $-$14.305 $\pm$ 0.036 & $-$14.625 $\pm$ 0.029 & $-$14.275 $\pm$ 0.028 & $-$14.426 $\pm$ 0.028 & $-$14.142 $\pm$ 0.027 & 3.931 $\pm$ 0.647 \\
\enddata
\tablecomments{\footnotesize Col. (1): The resolved $2 \times2$ pixel ($0\farcs7 \times 0\farcs7$) bins, ordered according to their radial distance from the peak of the nuclear SFR (the origin of coordinates in Figure~3: $\alpha = \rm 23^{h}03^{m}15\fs626$, $\delta = \rm 08\degree52\arcmin26\farcs16$). Cols. (2)--(6): Extinction-corrected flux of the molecular hydrogen rotational lines of H$_2$~$S(1)$ to H$_2$~$S(5)$; the uncertainty is the quadrature sum of the measurement uncertainty from MCMC fitting and the systematic uncertainty from the scaling of each spectrum. Col. (7): The power-law index $n$ derived from the five H$_2$ rotational lines, assuming that the H$_2$ rotational level populations are distributed as continuous-temperature components with $dN \propto T^{-n}dT$. (This table is available in its entirety in machine-readable form.)}
\label{tab:TableH2}
\end{deluxetable*}

\startlongtable
\setlength{\tabcolsep}{3pt}
\begin{deluxetable*}{ccccccccc}
\tabletypesize{\scriptsize}
\tablecolumns{9}
\tablecaption{Spatially Resolved Measurements of Velocity and Velocity Dispersion}
\tablehead{
\colhead{Region} & \colhead{$V_{\rm [Fe~{\small II}]}$} & \colhead{$\sigma_{\rm [Fe~{\small II}]}$} & \colhead{$V_{\rm [Ne~{\small II}]}$} & \colhead{$\sigma_{\rm [Ne~{\small II}]}$} & \colhead{$V_{\rm [Ne~{\small III}]}^{\rm narrow}$} & \colhead{$\sigma_{\rm [Ne~{\small III}]}^{\rm narrow}$} & \colhead{$V_{\rm [Ne~{\small III}]}^{\rm broad}$} & \colhead{$\sigma_{\rm [Ne~{\small III}]}^{\rm broad}$} \\
\colhead{} & \colhead{($\rm km\ s^{-1}$)} & \colhead{($\rm km\ s^{-1}$)} & \colhead{($\rm km\ s^{-1}$)} & \colhead{($\rm km\ s^{-1}$)} & \colhead{($\rm km\ s^{-1}$)} & \colhead{($\rm km\ s^{-1}$)} & \colhead{($\rm km\ s^{-1}$)} & \colhead{($\rm km\ s^{-1}$)}\\
\colhead{(1)} & \colhead{(2)} & \colhead{(3)} & \colhead{(4)} & \colhead{(5)} & \colhead{(6)} & \colhead{(7)} & \colhead{(8)} & \colhead{(9)}}
\startdata
NGC7469\_bin\_1 & 29 $\pm$ 2 & 96 $\pm$ 2 & 2 $\pm$ 3 & 98 $\pm$ 2 & 10 $\pm$ 1 & 119 $\pm$ 1 & $-$303 $\pm$ 10 & 415 $\pm$ 10 \\
NGC7469\_bin\_2 & 66 $\pm$ 1 & 103 $\pm$ 1 & 30 $\pm$ 1 & 103 $\pm$ 1 & 17 $\pm$ 1 & 119 $\pm$ 1 & $-$326 $\pm$ 9 & 418 $\pm$ 9 \\
NGC7469\_bin\_3 & 47 $\pm$ 2 & 109 $\pm$ 1 & 16 $\pm$ 2 & 104 $\pm$ 2 & 14 $\pm$ 1 & 120 $\pm$ 1 & $-$332 $\pm$ 8 & 464 $\pm$ 8 \\
NGC7469\_bin\_4 & 85 $\pm$ 1 & 113 $\pm$ 1 & 40 $\pm$ 1 & 108 $\pm$ 1 & 19 $\pm$ 1 & 121 $\pm$ 1 & $-$351 $\pm$ 8 & 485 $\pm$ 7 \\
NGC7469\_bin\_5 & 18 $\pm$ 2 & 78 $\pm$ 2 & $-$16 $\pm$ 3 & 78 $\pm$ 3 & 12 $\pm$ 1 & 111 $\pm$ 1 & $-$191 $\pm$ 11 & 314 $\pm$ 11 \\
NGC7469\_bin\_6 & 51 $\pm$ 1 & 86 $\pm$ 1 & 16 $\pm$ 1 & 85 $\pm$ 1 & 21 $\pm$ 1 & 113 $\pm$ 1 & $-$239 $\pm$ 11 & 330 $\pm$ 11 \\
NGC7469\_bin\_7 & 2 $\pm$ 2 & 74 $\pm$ 3 & $-$33 $\pm$ 5 & 73 $\pm$ 4 & 6 $\pm$ 1 & 112 $\pm$ 1 & $-$258 $\pm$ 7 & 408 $\pm$ 10 \\
NGC7469\_bin\_8 & 104 $\pm$ 1 & 95 $\pm$ 1 & 59 $\pm$ 2 & 92 $\pm$ 2 & 27 $\pm$ 1 & 117 $\pm$ 1 & $-$314 $\pm$ 9 & 438 $\pm$ 9 \\
NGC7469\_bin\_9 & 14 $\pm$ 3 & 86 $\pm$ 3 & $-$19 $\pm$ 4 & 81 $\pm$ 3 & 12 $\pm$ 1 & 113 $\pm$ 1 & $-$295 $\pm$ 8 & 420 $\pm$ 6 \\
NGC7469\_bin\_10 & 124 $\pm$ 1 & 104 $\pm$ 1 & 73 $\pm$ 2 & 100 $\pm$ 2 & 31 $\pm$ 1 & 120 $\pm$ 1 & $-$326 $\pm$ 7 & 477 $\pm$ 7 \\
\enddata
\tablecomments{\footnotesize Col. (1): The resolved $2 \times2$ pixel ($0\farcs7 \times 0\farcs7$) bins, ordered according to their radial distance from the peak of the nuclear SFR (the origin of coordinates in Figure~3: $\alpha = \rm 23^{h}03^{m}15\fs626$, $\delta = \rm 08\degree52\arcmin26\farcs16$). Cols. (2)--(9): The velocity and velocity dispersion of [Fe~{\scriptsize II}]~5.340\,$\mum$, [Ne~{\scriptsize II}]~12.814\,$\mum$, and the narrow and broad components of [Ne~{\scriptsize III}]~15.555\,$\mum$; the uncertainty only accounts for the standard deviation from MCMC fitting. The velocity dispersions have been corrected for instrumental broadening. (This table is available in its entirety in machine-readable form.)}
\label{tab:TableVel}
\end{deluxetable*}

\startlongtable
\setlength{\tabcolsep}{5pt}
\begin{deluxetable*}{cccccc}
\tabletypesize{\scriptsize}
\tablecolumns{6}
\tablecaption{Spatially Resolved Measurements of PAH Emission}
\tablehead{
\colhead{Region} & \colhead{log $f_{\rm PAH\,5-20}$} & \colhead{log $f_{\rm PAH\,6.2}$} & \colhead{log $f_{\rm PAH\,7.7}$} & \colhead{log $f_{\rm PAH\,11.3}$} & \colhead{log $\tau_{9.7}$} \\
\colhead{} & \colhead{($\rm erg~s^{-1}~cm^{-2}$)} & \colhead{($\rm erg~s^{-1}~cm^{-2}$)} & \colhead{($\rm erg~s^{-1}~cm^{-2}$)} & \colhead{($\rm erg~s^{-1}~cm^{-2}$)} & \colhead{}\\
\colhead{(1)} & \colhead{(2)} & \colhead{(3)} & \colhead{(4)} & \colhead{(5)} & \colhead{(6)}}
\startdata
NGC7469\_bin\_1 & $-$11.620 $\pm$ 0.041 & $-$12.745 $\pm$ 0.039 & $-$12.125 $\pm$ 0.039 & $-$12.486 $\pm$ 0.039 & $-$1.236 $\pm$ 0.098 \\
NGC7469\_bin\_2 & $-$11.651 $\pm$ 0.036 & $-$12.872 $\pm$ 0.031 & $-$12.165 $\pm$ 0.032 & $-$12.517 $\pm$ 0.032 & $-$6.552 $\pm$ 2.324 \\
NGC7469\_bin\_3 & $-$11.574 $\pm$ 0.029 & $-$12.635 $\pm$ 0.024 & $-$12.110 $\pm$ 0.024 & $-$12.435 $\pm$ 0.024 & $-$0.752 $\pm$ 0.032 \\
NGC7469\_bin\_4 & $-$11.615 $\pm$ 0.022 & $-$12.757 $\pm$ 0.008 & $-$12.178 $\pm$ 0.010 & $-$12.475 $\pm$ 0.010 & $-$5.584 $\pm$ 2.594 \\
NGC7469\_bin\_5 & $-$11.642 $\pm$ 0.056 & $-$12.737 $\pm$ 0.056 & $-$12.167 $\pm$ 0.056 & $-$12.627 $\pm$ 0.056 & $-$0.647 $\pm$ 0.022 \\
NGC7469\_bin\_6 & $-$11.651 $\pm$ 0.049 & $-$12.786 $\pm$ 0.049 & $-$12.158 $\pm$ 0.049 & $-$12.632 $\pm$ 0.049 & $-$6.606 $\pm$ 2.566 \\
NGC7469\_bin\_7 & $-$11.597 $\pm$ 0.036 & $-$12.637 $\pm$ 0.035 & $-$12.123 $\pm$ 0.036 & $-$12.499 $\pm$ 0.035 & $-$0.790 $\pm$ 0.031 \\
NGC7469\_bin\_8 & $-$11.664 $\pm$ 0.045 & $-$12.884 $\pm$ 0.043 & $-$12.161 $\pm$ 0.043 & $-$12.543 $\pm$ 0.043 & $-$6.771 $\pm$ 2.265 \\
NGC7469\_bin\_9 & $-$11.560 $\pm$ 0.035 & $-$12.574 $\pm$ 0.034 & $-$12.088 $\pm$ 0.034 & $-$12.438 $\pm$ 0.034 & $-$0.513 $\pm$ 0.021 \\
NGC7469\_bin\_10 & $-$11.670 $\pm$ 0.031 & $-$12.895 $\pm$ 0.028 & $-$12.198 $\pm$ 0.028 & $-$12.538 $\pm$ 0.028 & $-$1.026 $\pm$ 0.074 \\
\enddata
\tablecomments{\footnotesize Col. (1): The resolved $2 \times2$ pixel ($0\farcs7 \times 0\farcs7$) bins, ordered according to their radial distance from the peak of the nuclear SFR (the origin of coordinates in Figure~3: $\alpha = \rm 23^{h}03^{m}15\fs626$, $\delta = \rm 08\degree52\arcmin26\farcs16$). Cols. (2)--(5): Extinction-corrected flux of the integrated PAH emission from 5 to 20\,$\mum$, and the individual PAH features at 6.2\,$\mum$, 7.7\,$\mum$ (7.414\,$\mum$, 7.598\,$\mum$, and 7.85\,$\mum$), and 11.3\,$\mum$ (11.23\,$\mum$ and 11.33\,$\mum$); the uncertainty is the quadrature sum of the measurement uncertainty from MCMC fitting and the systematic uncertainty from the scaling of each spectrum. Col. (6): Total extinction at 9.7~$\mum$. (This table is available in its entirety in machine-readable form.)}
\label{tab:TablePAH}
\end{deluxetable*}

\appendix

\section{The PSF Models of the MIRI/MRS Observations}\label{sec:AppA}

\begin{figure}[t]
\figurenum{A1}
\center{\includegraphics[width=0.6\textwidth]{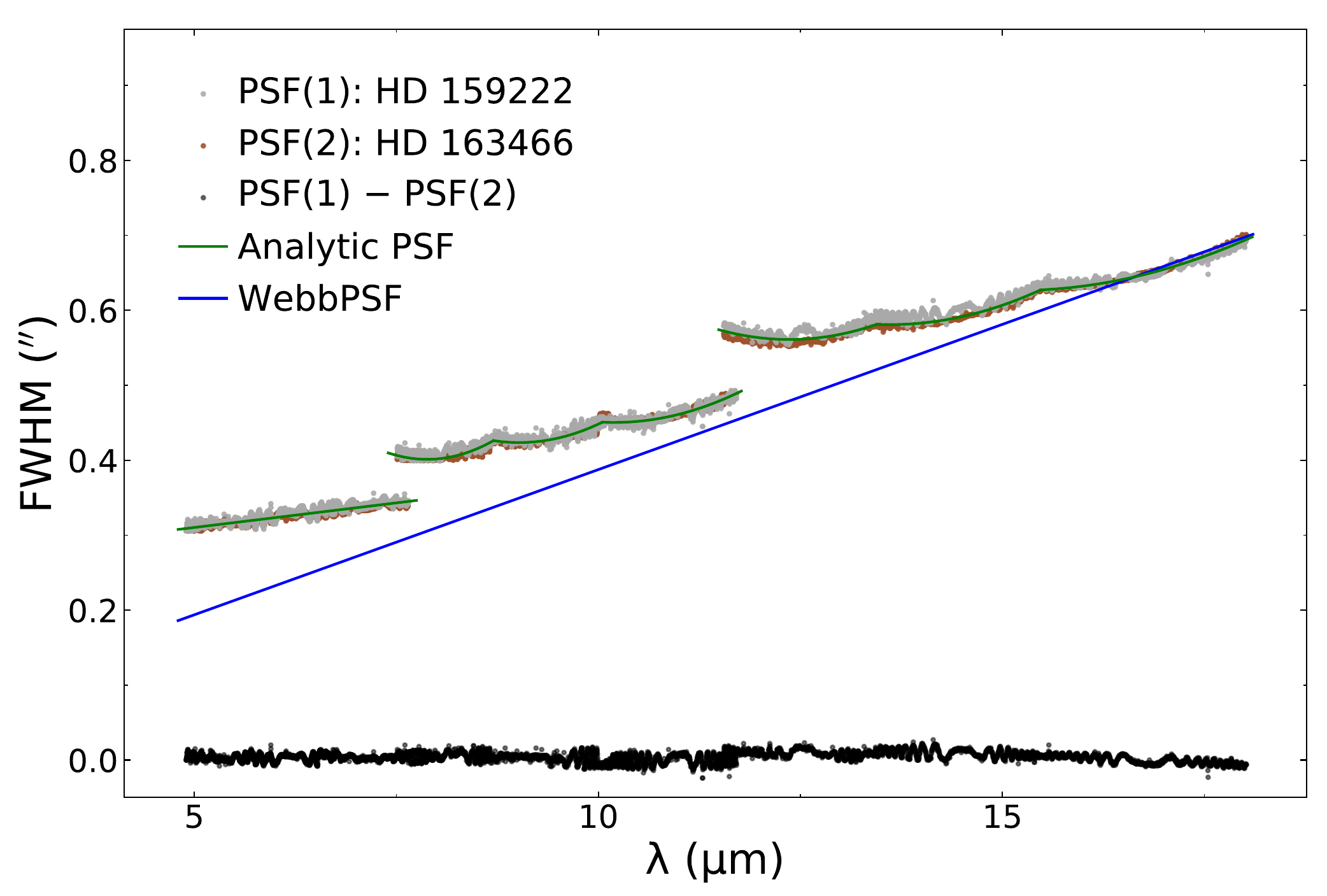}}
\caption{The FWHM as a function of wavelength of the PSF models for each slice within the first three channels of the MIRI/MRS data cubes. The PSF models of the stars HD~159222 (gray dots) and HD~163466 (brown dots) are measured from MIRI/MRS observations (ID 1050; PI: B. Vandenbussche) by fitting each slice of the data cubes with a two-dimensional Gaussian function. The black dots in the bottom show the difference between the FWHMs of the two sets of PSF models. The green curves represent the analytical fit of the FWHM as a function of wavelength based on the average FWHMs of the two sets of PSF models (Table~\ref{tab:psfs}). The blue line depicts the FWHM of the theoretical MIRI PSF model from {\tt WebbPSF}, ${\rm FWHM\ (\arcsec)} = 0.31\times\lambda\ (\mum)/8$, which applies to MIRI imaging observations.}
\label{fig:psf}
\end{figure}

We retrieve MIRI/MRS observations of two stars (HD~159222 and HD~163466) to model the PSFs, which are needed for convolving the data cubes to the same angular resolution (Section~\ref{section:sec2}). To obtain the FWHM of the PSF models as a function of slice number (wavelength), we fit each slice within the MIRI/MRS data cubes of the two stars with a two-dimensional Gaussian function. The two-dimensional Gaussian functions give only a first-order approximation of the PSFs of the MIRI/MRS data cubes; a more detailed investigation is warranted but beyond the scope of this Letter. Based on the two sets of PSF models (gray and brown dots in Figure~\ref{fig:psf}), we provide an analytical function to describe the variation of FWHM with wavelength for each channel (green curves in Figure~\ref{fig:psf}) by fitting the average FWHM of the two sets of PSF models with a linear or quadratic function. Table~\ref{tab:psfs} summarizes the best-fit parameters of the fitted functions. 

\begin{deluxetable*}{ccccccccc}
\tablenum{A1}
\tabletypesize{\scriptsize}
\tablecolumns{9}
\tablecaption{Analytical PSF Models for the MIRI/MRS Observations}
\tablehead{
\colhead{Band} &
\colhead{} & 
\colhead{Wavelength Range ($\mum$)} &
\colhead{} & 
\colhead{$\alpha$} &
\colhead{} & 
\colhead{$\beta$} &
\colhead{} & 
\colhead{$\gamma$} 
}
\startdata
Channels 1A--1C && $4.88-7.70$    && \nodata    && $0.013$  && $0.245$\\
\\
Channel 2A      && $7.47-8.70$    && $0.037$    && $-0.583$ && $2.698$\\
Channel 2B      && $8.70-10.05$   && $0.026$    && $-0.461$ && $2.503$\\
Channel 2C      && $10.05-11.77$  && $0.017$    && $-0.345$ && $2.207$\\
\\
Channel 3A      && $11.49-13.44$  && $0.017$    && $-0.425$ && $3.189$\\
Channel 3B      && $13.44-15.47$  && $0.013$    && $-0.360$ && $3.032$\\
Channel 3C      && $15.47-18.09$  && $0.008$    && $-0.251$ && $2.528$\\
\enddata
\tablecomments{\footnotesize The coefficients for ${\rm FWHM} = \alpha\lambda^{2}\,+\,\beta\lambda+\,\gamma$, with ${\rm FWHM}$ in units of arcseconds and $\lambda$ in units of $\mum$.}
\label{tab:psfs}
\end{deluxetable*}


\end{document}